\documentstyle[aps,epsf]{revtex}  

\begin{document}        

\baselineskip 14pt
\title{\vspace*{-3\baselineskip}
\hfill {\rm UR-1572}\\
\hfill {\rm ER-40685/933}\\
\hfill {\rm May 1999}\\
\vspace*{\baselineskip}
Top and gluons at lepton colliders\footnote{Presented by LHO 
at the 1999 Meeting of the Division of Particles and Fields of the APS,
Los Angeles, CA, Jan. 5--9, 1999.}}
\author{Cosmin Macesanu and Lynne H.~Orr}
\address{University of Rochester, Rochester, NY 14627-0171}
%
\maketitle              

\begin{abstract}        
In this talk we present results of an exact calculation of gluon
radiation in top production and decay at lepton colliders, including
all spin correlations and interferences.  We compare properties of 
gluons radiated in the production and decay stages and investigate
the sensitivity of interference effects to the top decay width.
\
\end{abstract}   	

\section{Introduction}               

Future high energy lepton colliders --- $e^+e^-$ and $\mu^+\mu^-$ will 
provide relatively clean 
environments in which to study top quark physics.  Although top production
cross sections are likely to be lower at these machines than at hadron
colliders, the color-singlet initial states and the fact that the 
laboratory and hard process center of mass frames coincide give 
lepton machines some advantages.  Strong interactions effects such as
those due to gluon radiation must still be considered, of course.  
Jets from radiated gluons can masquerade as quark jets, which can 
complicate top event identification and mass reconstruction from its
decay products, especially
for the hadronic decay modes.  

In this talk we consider gluon radiation in top quark production and 
decay (\cite{mo}; see also \cite{schmidt}).  We consider only  
collision energies well above the top pair 
production threshold, so that our results do not depend on whether 
the initial state consists of electrons or muons.
We focus on  distributions of radiated gluons and their effects
on top mass reconstruction.  

\begin{figure}[ht]	
\vskip -.75 cm
\hskip 1.5cm {\epsfxsize 4. truein \epsfbox{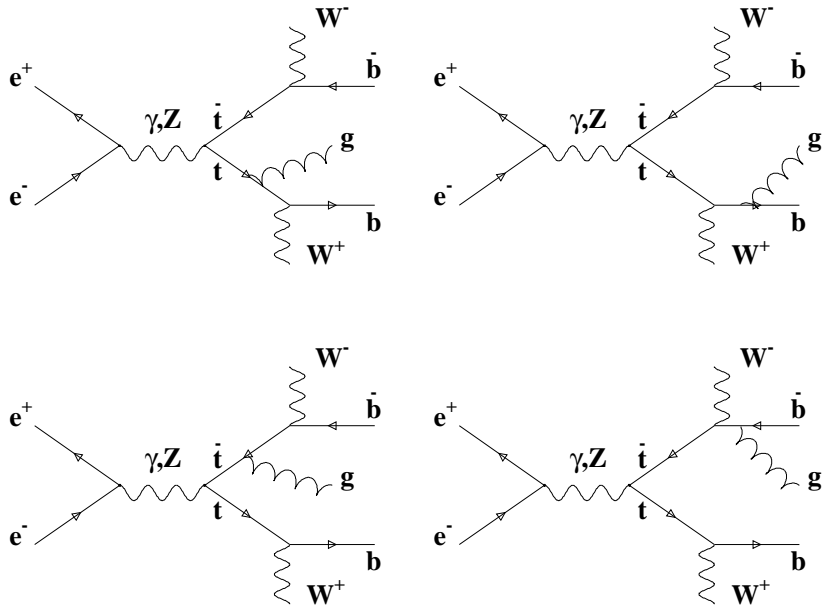}}   
\caption[]{
\label{diagrams}
\small Feynman diagrams for gluon emission in top production and decay at
lepton colliders.}
\end{figure}

In top events at lepton colliders, there are no gluons radiated from the 
color-singlet initial state.  Final-state gluon emission can occur in 
both the production and decay processes, with gluons emittied from the
top or bottom quarks (or antiquarks), as shown in Figure \ref{diagrams}.
Emission from the top quark contributes to both production- and decay-stage
radiation, depending on when the top quark goes on shell.  Emission 
from the $b$ quarks contributes to decay-stage radiation only.

\section{Monte Carlo Calculation}

The results presented here are from a Monte Carlo calculation of real
gluon emission in top quark production and decay: 
\begin{equation}
e^+e^- \rightarrow \gamma^*, Z^* \rightarrow t\bar{t} (g)
\rightarrow bW^+ \bar{b}W^-g\; .
\end{equation}
We compute the exact matrix elements for the diagrams shown in Figure 
\ref{diagrams} with all spin correlations and the bottom mass
included.  We keep the 
finite top width $\Gamma_t$ in the top quark propagator and include
all interferences between diagrams, and we use exact kinematics in
all parts of the calculation.
We do not include radiation from the decays of the $W$ boson;
this amounts to assuming either that the $W$ decays are leptonic or
that radiative hadronic $W$ decays can be identified and separated out,
for example by invariant mass cuts.

We are particularly interested in the reconstruction of the top quark
momentum (and hence its mass) from its decay products.  In an experiment this 
allows us both to identify top events and to measure the top quark's 
mass.  A complication arises when gluon radiation is present, because 
the emitted gluon may or may not be a top decay product.  If it is,
then we should include it in top reconstruction, i.e.\ we have 
$m_t^2=p_{Wbg}^2$ for decay-stage radiation.  But if the gluon
is part of top production, then we have $m_t^2=p_{Wbg}^2$.  It 
is therefore desirable to be able to identify and distinguish production-stage
gluons from those emitted in the decays.

Although this distinction cannot be made absolutely in an experiment, 
the various contributions can be separated in the calculation.  As noted
above,  gluon emission from the top quark (or antiquark) contributes
to both the production and decay stages.  These can be separated in the 
calculation as follows.  For definiteness, we consider gluon emission from
the top quark, shown in the bottom left diagram in Figure \ref{diagrams}.
The matrix element contains propagators for the top quark both before 
and after it radiates the gluon.  The matrix element therefore 
contains the factors 
\begin{equation}
ME \propto \left( {{1}\over{p_{Wbg}^2-m_t^2+im_t\Gamma_t}}\right)
\left( {{1}\over{p_{Wb}^2-m_t^2+im_t\Gamma_t}}\right)\; .
\end{equation}
The right-hand side can be rewritten as
\begin{equation}
{{1}\over{2p_{Wb}*p_{Wbg}}}
\left( {{1}\over{p_{Wb}^2-m_t^2+im_t\Gamma_t}} -
{{1}\over{p_{Wbg}^2-m_t^2+im_t\Gamma_t}}\right)\; .
\end{equation}
This separates the production and decay contributions to the matrix element 
because the 
two terms in parentheses  peak respectively at $p_{Wb}^2=m_t^2$ (production
emission) and $p_{Wbg}^2=m_t^2$ (decay emission).  The cross section in 
turn contains separate production and decay contributions.  It also contains
interference terms, which in principle confound the separation but in 
practice are quite small.

In fact the interference terms are  interesting in their own right, although
not for top reconstruction.  In particular, the interference between 
production- and decay-stage radiation can be sensitive to the top
quark width $\Gamma_t$, which is about 1.5 GeV in the Standard Model.\cite{kos}
The interference between the two propagators shown above can be 
thought of as giving rise to two overlapping Breit-Wigner resonances.  The
peaks are separated roughly by the gluon energy, and each curve
has width $\Gamma_t$.  Therefore when the gluon energy becomes comparable 
to the top
width, the two Breit-Wigners overlap and interference can be substantial.
In constrast, if the gluon energy is much larger than $\Gamma_t$, overlap
and hence interference is negligible.  Hence the amount of interference serves
as a measure of the top width.  We will explore this more below.

\section{Numerical Results}

\subsection{Overall Gluon Properties}

\begin{figure}[ht]	
\centerline{\epsfxsize 4.1 truein \epsfbox{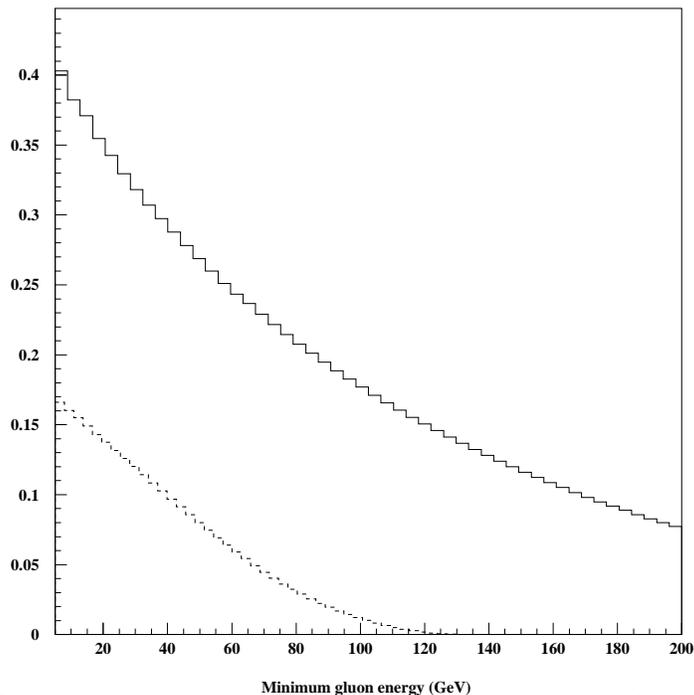}}   
\vskip -.25 cm
\caption[]{
\label{prodfrac}
\small The fraction of gluon emissions radiated in the production stage, as
a function of minimu gluon energy, for center-of-mass energy 1 TeV (solid
line) and 500 GeV (dashed line).}
\end{figure}

We begin our numerical results with the  relative contributions of production-
and decay-stage radiation to the total cross section.  Figure \ref{prodfrac}
shows the fraction of the total cross section due to production stage 
emission, in events with an extra gluon.  The solid line is for center-of-mass
energy 1 TeV, and the dashed line is for 500 GeV.  Both curves fall off 
as the minimum gluon energy increases; this reflects the decrease in 
phase space for emitted gluons.  The production fraction is higher at 
a 1 TeV collision energy than at 500 GeV --- again this reflects 
phase space --- but decay-stage radiation always dominates for both cases.

\begin{figure}[ht]	
\centerline{\epsfxsize 3.5 truein \epsfbox{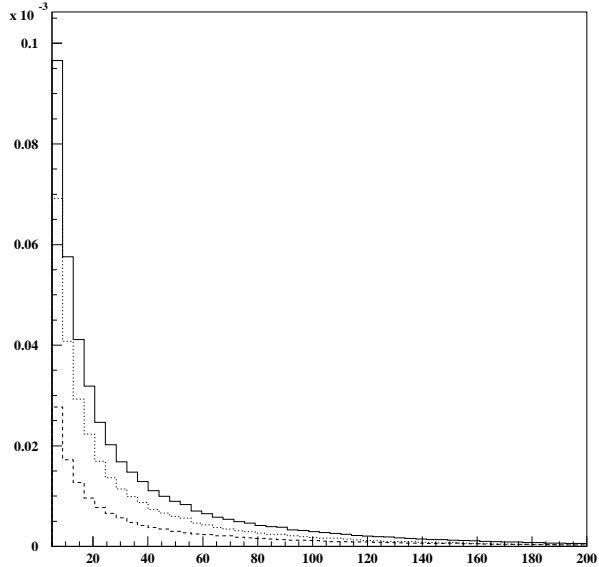}}   
\vskip -.5 cm
\caption[]{
\label{energy}
\small The  spectrum of radiated gluons as a function of gluon energy in
GeV.  Dashed histogram:  production-stage radiation.  Dotted histogram: 
decay-stage radiation.  Solid histogram:  total.}
\end{figure}

Figure \ref{energy} shows the total gluon energy spectrum for an intermediate
collision energy of 750 GeV along with its decomposition into production
(dashed histogram) and decay (dotted histogram) contributions.  Again we see 
that deca-stage radiation dominates.  Otherwise the spectra are not 
very different; both exhibit the rise at low energies due to the 
infrared singularity characteristic of gluon emission, and both fall
off at high energies as phase space runs out.

\subsection{Mass Reconstruction}

\begin{figure}[ht]	
\centerline{\epsfxsize 3.5 truein \epsfbox{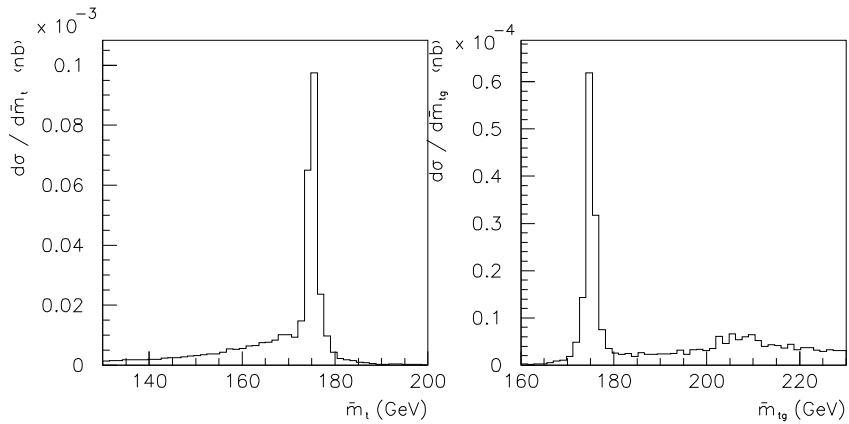}}   
\vskip -.2 cm
\caption[]{
\label{masses}
\small The  top invariant mass spectrum without (left) and with (right) 
the gluon momentum included, for center-of-mass energy 600 GeV.}
\end{figure}

We now turn to the question of mass reconstruction.  
Figure \ref{masses} shows top invariant mass distributions with and without
the extra gluon included. In both cases there is a clear peak  at the 
correct value of $m_t$.  In the left-hand plot, where the gluon
is not included in the reconstruction, we see a low-side tail due to 
events where the gluon was radiated in the decay.  Similarly, 
in the right-hand plot we see a high-side tail due to events where the 
gluon was radiated in association with production, and was included when it 
should not have been.  

\begin{figure}[ht]	
\centerline{\epsfxsize 3.5 truein \epsfbox{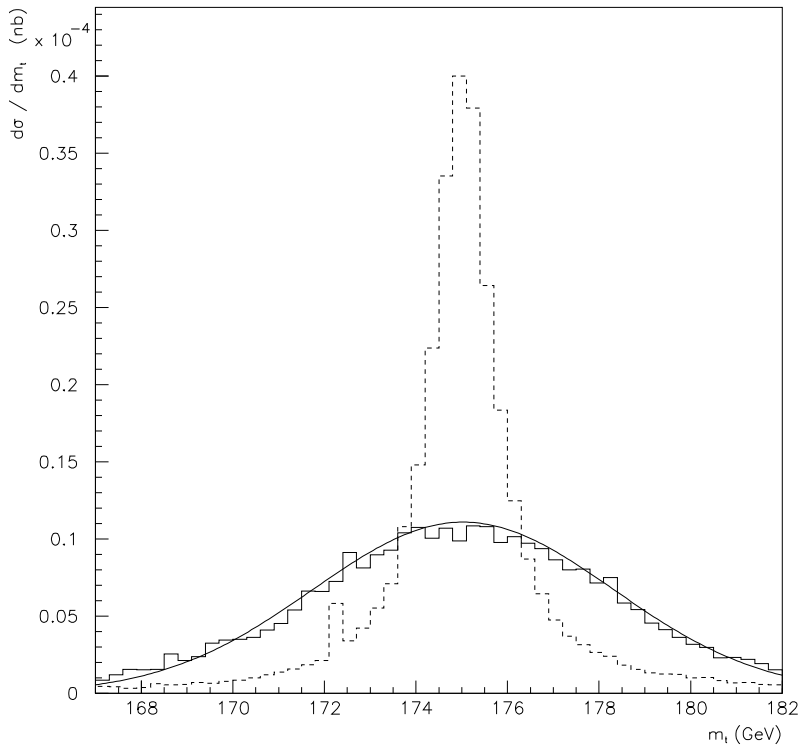}}   
\vskip -.2 cm
\caption[]{
\label{masssmear}
\small The  top invariant mass spectrum with $b$-gluon angle selection
criteria (dotted histogram), for center-of-mass energy 600 GeV and minimum gluon
energy 10 GeV.  The solid curve and histogram show the effects of energy
smearing.}
\end{figure}

The narrowness of the peaks and the length of the tails in Figure \ref{masses}
suggests that an invariant mass cut would be useful to separate the two
types of events.  We can do even better by considering cuts on the angle 
between the gluon and the $b$ quarks.    This works because 
although there is no collinear singularity for radiation from massive
quarks, the distribution of gluons radiated from $b$ quarks peaks
close to the $b$ direction.  Such gluons are emitted in decays.
The dotted histogram in Figure \ref{masssmear} shows the top
mass distribution that results from using proximity of the gluon to the 
$b$ quarks to assign the gluon.  

Of course an importnat reason the cuts are so effective is that we work at the 
parton level.  The experimentalists do not have that luxury, and, as one
would expect, hadronization and detector effects are likely to cloud
the picture.  The solid histogram in Figure \ref{masssmear} shows the
mass distribution after including energy smearing; the solid curve is a 
Gaussian fit.   
The spread in the measured energies is parametrized by Gaussians with
widths $\sigma=0.4 \sqrt{E}$ for quarks and gluon, and $ 
\sigma=0.15 \sqrt{E}$ for the $W$'s.  We see that the central value does not 
shift, but the distribution is significantly wider.

\subsection{Interference and Sensitivity to $\Gamma_t$}

Finally, we return to the subject of interference.  As mentioned above, the 
interference between the production- and decay-stage radiation
is sensitive to the total width of the top quark $\Gamma_t$.  
However because the interference is in general small, we need to find
regions of phase space where it is enhanced.  This question was considered
in Ref.~\cite{kos} in the soft gluon approximation, where it was found that 
the interference was enhanced when there was a large angular separation
between the $t$ quarks and their daughter $b$'s.

\begin{figure}[ht]	
\centerline{\epsfxsize 3.5 truein \epsfbox{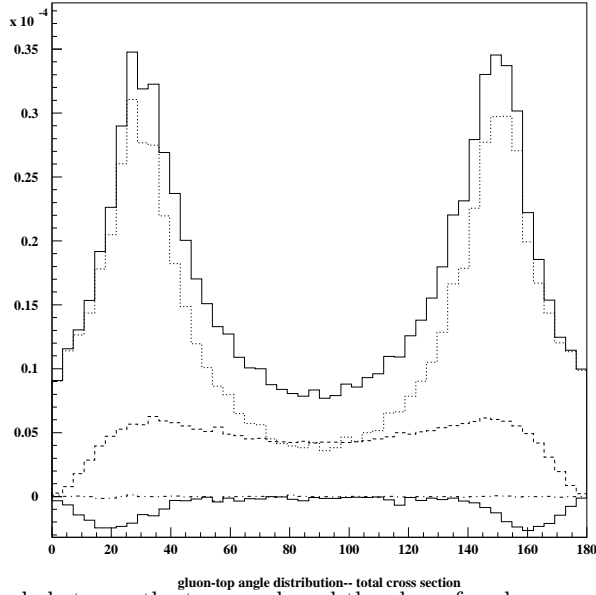}}   
\vskip -.2 cm
\caption[]{
\label{thetagt}
\small The  distribution in angle between the top quark and the gluon for 
gluon energies from 5 to 10 GeV, $\cos\theta_{tb}<0.9$, and 750 GeV collision
energy.  The upper solid histogram is the  total and the other histograms
represent the individual contributions:  dotted:  decay; dashed: production;
dot-dashed: decay-decay interference; solid:  production-decay interference.}
\end{figure}

\begin{figure}[ht]	
\centerline{\epsfxsize 3.5 truein \epsfbox{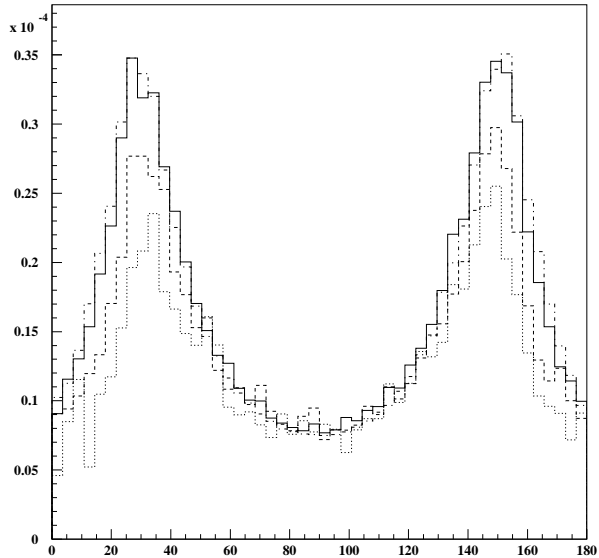}}   
\vskip -.2 cm
\caption[]{
\label{width}
\small The  distribution in angle between the top quark and the gluon for 
gluon energies from 5 to 10 GeV, $\cos\theta_{tb}<0.9$, and 750 GeV collision
energy.  The histograms correspond to different values of the top 
width $\Gamma_t$:  dot-dashed: 0.1 GeV; solid: 1.5 GeV (SM); dotted: 5. GeV;
dotted: 20 GeV.}
\end{figure}

Here we examine whether
the result of \cite{kos} survives the exact calculation.  Figure \ref{thetagt}
shows that it does.  There we plot the distribution in the angle between the 
emitted gluon and the top quark for gluon energies between 5 and 10 
GeV and with $\cos\theta_{tb}<0.9$.  The center-of-mass energy is 750 GeV.
The  histograms show the decomposition into the various contributions.  
The negative solid histogram is the production-decay interference, and 
we see that not only is it substantial, it is also destructive.  That means
that the interference serves to suppress the cross section.  If the 
top width is increased, the interference is larger, further suppressing the 
cross section.  this is illustrated in Figure \ref{width}, which
shows the cross section for different values of the top width.  Although
the sensitivity does not suggest a precision measurement, it is worth
noting that the top width is difficult to measure by any means, and it
is the total width that appears here.  At the very least such a measurement 
would serve as a consistency check.

\section{Conclusion}

In summary, we have presented preliminary results from an exact
parton-level calculation of real gluon radiation in top production and decay
at lepton colliders, with the $b$ quark mass and finite top width, as well
as all spin correlations and interferences included.  We have indicated
some of the issues associated with this gluon radiation in top mass
reconstruction and top width sensitivity in the gluon distribution.  Further
work is in progress.

This work was supported in part by the U.S. Department of Energy and the 
National Science Foundation.

\end{document}